# Early *in vivo* Radiation Damage Quantification for Pediatric Craniospinal Irradiation Using Longitudinal MRI for Intensity Modulated Proton Therapy

Chih-Wei Chang, PhD[1], Matt Goette, PhD[1], Nadja Kadom, MD[2], Yinan Wang, MD[1], Jacob Wynne, MD[1], Tonghe Wang, PhD[3], Tian Liu, PhD[4], Natia Esiashvili, MD[1], Jun Zhou, PhD[1], Bree R. Eaton, MD[1*] and Xiaofeng Yang, PhD[1*]

[1]Department of Radiation Oncology and Winship Cancer Institute, Emory University, Atlanta, GA 30308

[2]Department of Radiology and Imaging Sciences, Emory University and Children's Healthcare of Atlanta, Atlanta, GA 30308

[3]Department of Medical Physics, Memorial Sloan Kettering Cancer Center, New York, NY, 10065

[4]Department of Radiation Oncology, Mount Sinai Medical Center, New York, NY 10029

Email: xiaofeng.yang@emory.edu and brupper@emory.edu

**Running title:** Quantification of in vivo proton damage for pediatric craniospinal irradiation
**Manuscript Type:** Original Research

**Abstract**

**Purpose:** Proton vertebral body sparing craniospinal irradiation (VBS CSI) treats the thecal sac while avoiding the anterior vertebral bodies in effort to reduce myelosuppression and growth inhibition. However, robust treatment planning needs to compensate proton range uncertainty, contributing unwanted doses within the vertebral bodies. This work aims to develop an early *in vivo* radiation damage quantification method using longitudinal magnetic resonance (MR) scans to quantify dose effect during fractionated CSI.

**Materials and methods:** Ten pediatric patients were enrolled in a prospective clinical trial of proton VBS CSI receiving 23.4-36 Gy. Monte Carlo robust planning was used with spinal clinical target volumes defined as the thecal sac and neural foramina. T1/T2-weighted MR scans were acquired before, during, and after treatments to detect transition from hematopoietic to less metabolically active fatty marrow. MR signal intensity histograms at each time point were analyzed and fitted by multi-Gaussian models to quantify radiation damages.

**Results:** Fatty marrow filtration was observed on MR images as early as the fifth fraction of treatment. Maximum radiation-induced marrow damage occurred 40-50 days from the treatment start, followed by marrow regeneration. The mean damage ratios were 0.23, 0.41, 0.59, and 0.54 corresponding to 10, 20, 40, and 60 days from the treatment start.

**Conclusions:** We demonstrated a non-invasive method to identify early vertebral marrow damage based on radiation-induced fatty marrow replacement. The proposed method can be potentially used to quantify the quality of CSI vertebral sparing to preserve metabolically active hematopoietic bone marrow.

# 1    Introduction

Craniospinal irradiation (CSI) is a curative treatment for several pediatric central nervous system malignancies. The CSI target volume includes the entire brain and thecal sac as the clinical target volume to minimize the risk of tumor dissemination throughout the neuroaxis [1]. Conventional photon CSI treatment may induce various short- and long-term side effects, including odynophagia, anorexia, bone marrow suppression causing lymphopenia, and secondary malignant neoplasms [2-5] in these young patients with very favorable prognoses. Clinical evidence also has shown that growing children can further develop spinal lordosis, kyphosis, or scoliosis when growth plates are asymmetrically irradiated [6-8]. Consequently, pediatric radiation oncologists have historically recommended treating vertebral bodies and growth plates wholistically for growing children [9]. Even with this practice, Paulino *et al.* [10] reported that 16.7% and 54.5% of patients developed scoliosis 15 years after their vertebral ossification centers received CSI doses of 18 - 24 Gy and 34.2 - 40 Gy, respectively. Thus, a treatment technique with conformal dose delivery and avoidance of the vertebral bodies is desired for CSI in children to reduce toxicity and improve quality of life.

Intensity modulated proton therapy [11] is an attractive technique for the treatment of pediatric CNS malignancies due to its superior dose conformality and reduced integral dose to surrounding healthy tissues. Clinical evidence has shown that proton CSI can reduce radiation toxicity and achieve equivalent long-term disease control relative to photon therapy [12-15]. Unlike photon beams, the integral depth dose curve of proton beams exhibits a distal peak near the end of the proton range (Bragg peak) that enables dose deposition without significant exit dose. Proton CSI has the potential to leverage this physical feature to treat the entire thecal sac while sparing the spinal growth plate. However, current proton therapy treatment planning usually includes a margin of 3.5% reserved for proton range uncertainty [16-18]. Such uncertainty requires the inclusion of parts of the vertebral bodies during treatment planning to ensure that the target volume receives adequate dose coverage, and the sparing of growth plates can be compromised.

Given this uncertainty in proton therapy, a method of quantifying the *in vivo* proton damage during treatments would be valuable to verify the accuracy of treatments and to facilitate adaptive re-planning, if necessary, when using a steep dose gradient for vertebral body sparing (VBS). After proton CSI, radiation-induced fatty marrow infiltration can be observed in the spine following treatment using magnetic resonance imaging (MRI) [19, 20]. Replacement of hematopoietic marrow provides physiologic evidence to retrospectively investigate the potential *in vivo* proton range uncertainty [21, 22]. However, there remains a paucity of data from which to deduce when marrow conversion happens and how to use this information to protect vertebral growth plates.

Although fatty marrow replacement has been detected at the end of treatment [23, 24], whether fatty marrow may be observed on MR images between earlier treatment fractions in children remains unknown. This study aims to perform MRI at specified intervals during proton VBS CSI to detect how early radiographic marrow changes become evident and to evaluate whether the planned radiation dose deposition in bone can be correlated to proportional proton damage within vertebral marrow. These findings may potentially be used to support real-time medical decision-making, for instance to determine if growth plates are sufficiently spared or replanning is necessary to reduce excessively conservative proton range margins. Quantification of uncertainty will be considered to demonstrate the reliability, applicability, and feasibility of the proposed method for CSI intensity modulated proton therapy with vertebral body sparing.

## 2 Materials and methods

*Patient identification, scan characteristics, and treatment planning*

Patients were enrolled in a prospective clinical trial of proton VBS CSI in children (NCT04276194) with any malignancy and ages less than or equal to 18-year-old. Patients and their families signed formal consent for this IRB approved protocol. Details of the protocol specified treatment and primary outcomes are separately reported. This work focuses on the secondary endpoint of evaluating the feasibility of using MRI performed during CSI as a method of *in vivo* quantification of radiation dose deposition to bone marrow.

Patients underwent CT simulation in the supine position using a five-point thermoplastic mask for immobilization. CT scans were acquired from the vertex to pelvis using a Siemens SOMATOM Definition Edge scanner with resolution 0.98x0.98x1.0 mm$^3$. Prescribed doses ranged from 15-36 Gy (RBE) in 1.5-1.8 Gy (RBE) daily fractions, assuming a relative biological effectiveness (RBE) of 1.1 per IAEA/ICRU guidelines [25, 26]. The clinical target volume (CTV) included the entire cranial contents and thecal sac surrounding the spinal cord and nerve roots. The cranium was treated with a single posteroanterior field, or two posterior oblique fields and the spinal thecal sac was treated with 1-2 posteroanterior fields. Treatment was planned in RayStation using Monte Carlo (MC) robust optimization with 5 mm positional and 3.5% range uncertainties.

MRI scans were acquired at baseline (within 6 weeks prior to starting CSI), during CSI treatment at approximately fraction 7, 13, and 20, and 4-week following completion of treatment. A window of ±3 days was allowed for MRI scans during CSI treatment. MRI sequences included T1- and T2-weighted turbo spin echo MR sequences without contrast acquired in the sagittal plane on a Siemens MAGNETOM Aera 1.5T scanner with slice spacing 0.78-4 mm and pixel spacing 0.78x0.78-1.09x1.09 mm$^2$. Echo time and repetition time ranges were 10-13 ms and 550-771 ms for T1-weighted images and 69-102 ms and 1500-4420 ms for T2-weighted images, respectively. Table 1 summarizes patient demographics and diagnosis as well as imaging and treatment details.

**Table 1.** Summary of the identified pediatric patients.

| Patient | Age | Diagnosis | CSI spine dose (Gy) | Dose per fraction (Gy) | MR scans |
|---|---|---|---|---|---|
| 1 | 3.8 | Atypical teratoid rhabdoid tumor | 36 | 1.8 | 6 |
| 2 | 9.3 | Medulloblastoma | 36 | 1.8 | 6 |
| 3 | 7.1 | Medulloblastoma | 36 | 1.8 | 6 |
| 4 | 16.6 | NGGCT | 36 | 1.8 | 2 |
| 5 | 15.0 | NGGCT | 36 | 1.8 | 2 |
| 6 | 3.2 | Medulloblastoma | 36 | 1.8 | 6 |
| 7 | 11.1 | NGGCT | 36 | 1.8 | 6 |
| 8 | 10.5 | Acute myeloid leukemia | 15 | 1.5 | 4 |
| 9 | 5.7 | Pineoblastoma | 36 | 1.8 | 5 |
| 10 | 8.5 | Medulloblastoma | 23.4 | 1.8 | 2 |

MRI scans during treatment were reviewed by a pediatric neuro-radiologist and compared to the baseline image to determine at which earliest time point RT induced bone marrow changes could be detected visually. Quantitative image processing was also performed to evaluate RT induce bone marrow change, as described below.

*Imaging processing*

Varian Velocity™ was used to register MRI (moving images) to CT (target images). Lumbar spine contours from the level of L1-L5 were propagated from the CT images to MRI. The lumbar vertebrae were selected for their relatively fixed position relative to other vertebrae, which reduces registration error between imaging modalities. Two-Gaussian distribution sum models were used to analyze MRI intensity within the lumbar contours. Before treatment, bone marrow was intact and, as a result, the MR intensity histogram was reasonably approximated by a single Gaussian distribution. During treatment, irradiated hematopoietic bone marrow is converted to fatty marrow. The two marrow types exhibit distinct MR signal intensity such that their intensity histograms show two separate Gaussian distributions. Figure 1 shows an example of a MR histogram as a probability distribution after spine irradiation. Eq. (1) defines the radiation damage ratio by integrating the area under the total MRI signal distribution curve ($G(x)$) from the intersection point ($x_{int}$) of the two intensity distributions. $x$ is the relative MRI signal intensity. Relative MRI signal is used for analysis to reduce MRI signal discrepancies from each patient. Raw MRI signal was divided by the mean from the last 5% of the raw signal for each scan from the intensity histogram since those values were minimally impacted by irradiation.

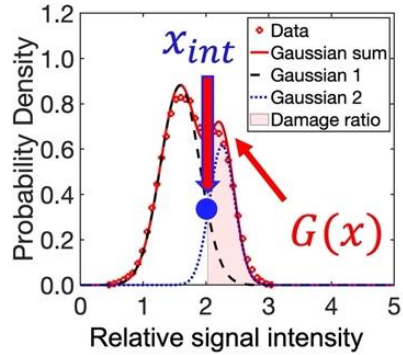

**Figure 1.** MR histogram presented as a probability density distribution where $x_{int}$ is the intersection point between Gaussian 1 and Gaussian 2 and $G(x)$ is the total distribution (Gaussian sum).

$$Damage\ ratio = \int_{x_{int}}^{\infty} G(x)dx \quad (1)$$

## 3 Results

*Damage ratio variation along with D1 dose and time*

We evaluated radiation-induced fatty marrow infiltration within the L1-L5 vertebrae for ten patients at using multiple MRIs acquired throughout treatment. Figures 2(a1)-(a2) depict the proton isodose profile across the region of interest. Of note, parts of the vertebrae receive the full prescription dose of 36 Gy (RBE) for patient 1. Figures 2(b1)-(b2) display the pre-treatment T1-weighted MRI and corresponding signal intensity distribution within the lumbar vertebrae as a Gaussian distribution. After treatment begins, fatty marrow replacement is observed in both MRI and CT-MRI fusion images, shown in Figures 2(c1)-(f1) and Figures 2(c2)-(f2). A dose-response relationship was observed for fatty marrow conversion. Figure 2(c3) demonstrates that relative MR signal intensity increases after spine irradiation with a distribution that deviates from a single Gaussian; instead, the new distribution is better approximated by a Gaussian sum composed of two Gaussian functions. Figure 2(c3) also displays the damaged area (red shadow), which begins at the intersection of the overlapping Gaussian models. By integrating the red area under the Gaussian sum in Figure 2(c3), we find a damage ratio of 0.446 at the 8$^{th}$ fractional dose. Figures 2(d3)/(e3)/(f3) illustrate the relative MR intensity distributions and damage ratios at different treatment stages.

Figure 3 shows the treatment plan for patient 8. Because full prescription dose was delivered to portions of the L1-L5 vertebral bodies, fatty marrow conversion was detected in T2-weighted MRIs, displayed in Figure 3(c1)-(f1). CT-MR fusion images in Figure 3 depict MR intensity migration with increased dose. Figure 3(c3)-(e3) reveals damage ratios quantified from MR images at different time.

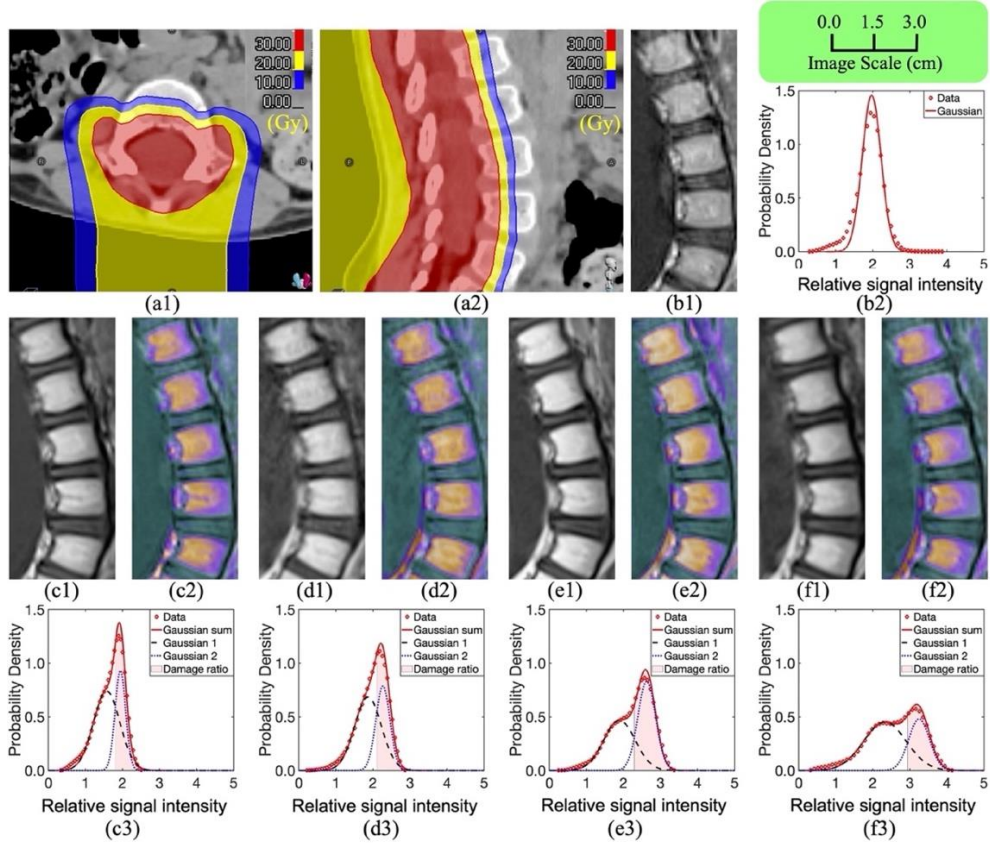

**Figure 2.** Demonstration of radiation-induced fatty infiltration at the lumbar spine, the L1-L5 level, from patient 1 with the prescribed CSI spine dose of 36 Gy in 20 fractions. The radiotherapy treatment planning CT with MC dose for (a1) transversal image, L3 level, and (a2) sagittal image. (b1) Sagittal T1-weighted MRI image and (b2) distribution of MRI intensity within vertebral bodies, the L1-L5 level, acquired 13 days before the treatment start date. Sagittal T1-weighted MRI images ((c1)/(d1)/(e1)/(f1)), CT-MRI fusion images ((c2)/(d2)/(e2)/(f2)), and distribution of MRI intensity within vertebral bodies, L1-L5 level ((c3)/(d3)/(e3)/(f3)), acquired (c1)-(c3) 15 days (8$^{th}$ fraction) after, (d1)-(d3) 26 days (15$^{th}$ fraction) after, (e1)-(e3) 37 days after, and (f1)-(f3) 45 days after the CSI treatment start date. The damage ratios are given by (c3) 0.446, (d3) 0.459, (e3) 0.550, and (f3) 0.392.

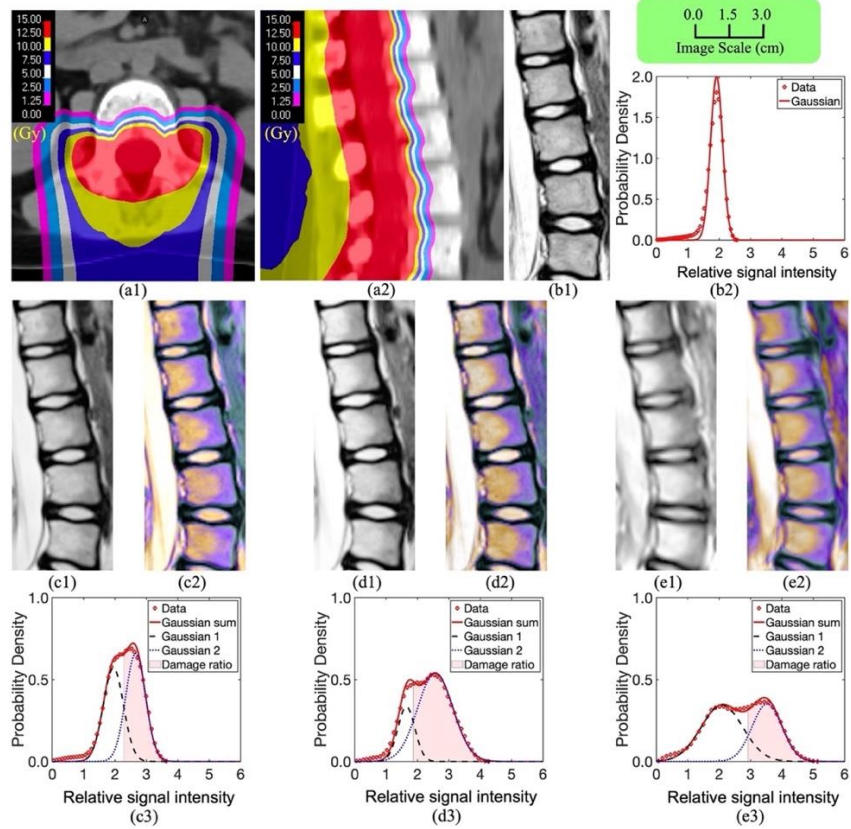

**Figure 3.** Demonstration of radiation-induced fatty marrow infiltration within the L1–L5 lumbar spine from patient 8 with the prescribed CSI spine dose of 15 Gy in 10 fractions. Radiotherapy treatment planning CT is presented with MC dose in (a1) the transverse plane at the L3 level and (a2) in the sagittal plane. (b1) Sagittal T2-weighted MRI and (b2) distribution of MRI intensity within the L1 – L5 vertebral bodies, acquired 25 days before treatment start. Sagittal T2-weighted MRI images ((c1)/(d1)/(e1)), CT-MRI fusion images ((c2)/(d2)/(e2)), and distribution of MRI intensity within the L1 – L5 vertebral bodies ((c3)/(d3)/(e3)), acquired (c1)-(c3) 11 days (9$^{th}$ fraction) after, (d1)-(d3) 18 days after, and (e1)-(e3) 65 days after the CSI treatment start date. The damage ratios are given by (c3) 0.536, (d3) 0.738, and (e3) 0.439.

Ultimately, we included data from all patients' MRIs to infer the correlation for the damage ratio variation based on the dose to 1% of the volume (D1) and time. The D1 dose was selected as a figure of merit because this quantity was sensitive to fractional doses. Figure 4(a) depicts a positive correlation between the damage ratio and D1 dose. Figure 4(b) shows a quadratic relation between the damage ratio and time with a concave profile, which suggests marrow regeneration following maximal damage.

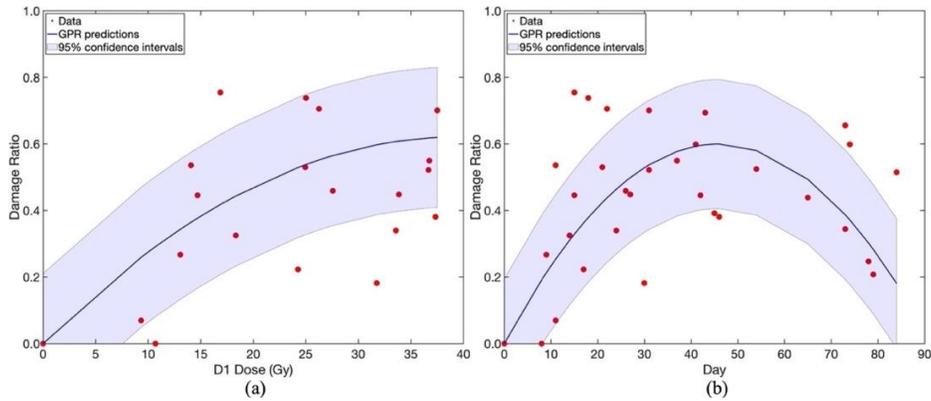

**Figure 4.** Damage ratio variation along with (a) D1 doses and (b) time. Blue lines are predicted by Gaussian process regression while blue shadows represent 95% confidence intervals.

*Dose-signal intensity curve*

Figures 5 show registered MR images exhibiting fatty marrow infiltration at different time with different accumulative doses. Only hematopoietic marrow can be observed before treatments. Fatty marrow starts to form only after proton irradiation. As proton accumulative dose increases, the region of fatty marrow increases and the edge of the area aligns well with overlaid isodose lines, which correspond to the distal falloff doses in the vertebra bodies. Figure 6 depicts the dose-response relationship as a function of MR signal intensities. An approximate threshold of 2.25 Gy is observed in these data and doses increase rapidly with signal intensities. Ultimately, the curve saturates at 34-35 Gy.

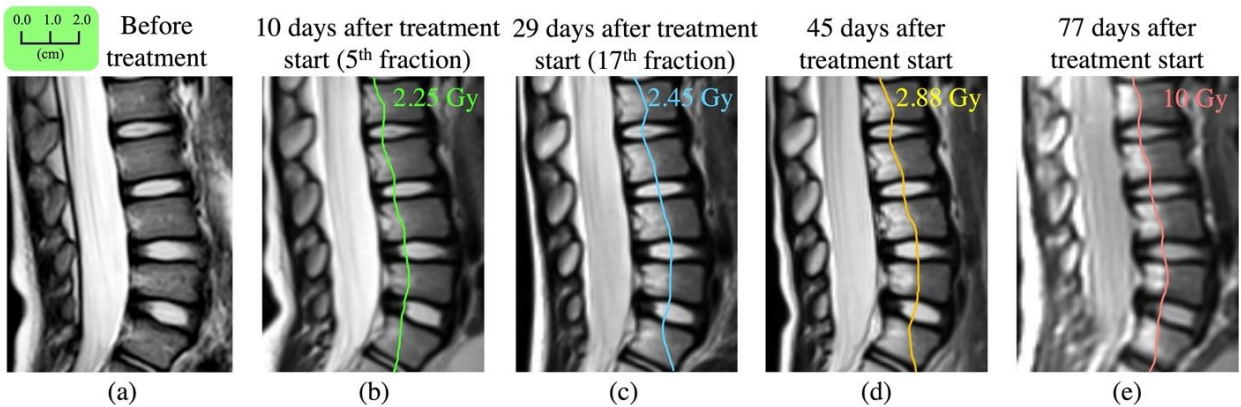

**Figure 5.** Demonstration of radiation-induced fatty marrow infiltration within the lumbar spine from patient 9. MR images with isodose lines were acquired (a) 42 days before, (b) 10 days after, (c) 29 days after, (d) 45 days after, and (e) 77 days after treatment start.

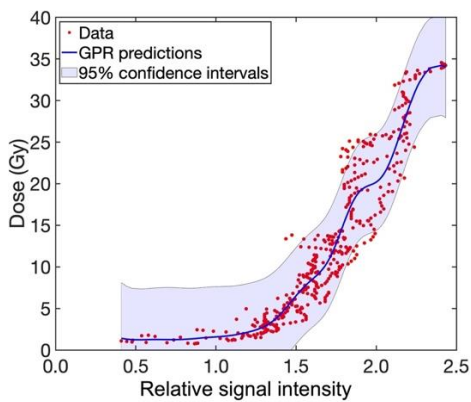

**Figure 6.** Dose-signal intensity curve (blue line) fitted by Gaussian process regression using the anteroposterior data points within the vertebra depicted in Figure 4(b1)-(b5). The blue shadow shows the 95% confidence interval.

## 4 Discussion

While fatty marrow replacement has previously been reported following treatment [21], we use multiple MR scans acquired during treatment to demonstrate an early dose-effect relationship. We observe the correlation between radiation-induced fatty marrow infiltration and D1 doses and isodose lines can be found to align to the edge of fatty marrow in vertebral bodies during inter-fraction treatment.

Figure 4(a) reveals a steep dose-effect relationship between D1 doses of 15 and 30 Gy, with a decrease in relative effect thereafter, possibly due to a maximum of accumulated damage acquired during previous fractions. Figure 6 also demonstrates a sharp inflection in MR signal intensity observed between 2.25 and 34 Gy, with a shoulder region between 34 and 35 Gy (RBE). These observations are consistent with previously published findings that an existence of a steep dose-effect correlation in a 15-35-Gy range [27-29]. Figure 4(b) suggests damage ratios reach a maximum between 40 and 50 days from the start of treatment with a subsequent decrease, likely representing marrow regeneration as observed in Figure 2(f1)-(f2) and corresponding to that reported in a previous study [24]. Because MR data are lacking for cumulative doses of 0–10 Gy, Figure 4 includes an extrapolation region across this range. MR scans acquired with each treatment fraction may reduce uncertainty and further bolster dose-effect correspondence.

Figure 5(b)-(d) depict that the distal falloff doses cause the fatty marrow replacement in the vertebral bodies and isodose lines can be found to align the edges of fatty marrow. Although the accumulative doses are different in Figure 5(b)-(d), the quantities of each isodose line do not change significantly because those isodose lines belongs to distal falloff of the prescription dose regions. Figure 5(e) shows the decrease in fatty marrow due to potential marrow regeneration such that the marrow edge is pulled back and the edge aligns to a high dose isodose line. The marrow edge aligns well to the 10-Gy isodose agreeing to the literature [30] using MR images acquired more than two months from treatment start. However, Figure 5(b)-(d) depict that marrow edges match isodose lines ranging from 2.25-2.88 Gy. The results suggest that MR images acquired during treatments could be used for building dose-and-signal-intensity correlations to avoid biases due to marrow regeneration.

A previous dosimetric study has confirmed fatty marrow conversion with doses as low as 16 Gy [31]. Here, we demonstrate that marrow conversion is observed on MRI with doses as low as 2.25 Gy (Figure 6). Partition growth inhibition and hypoplasia of paraspinal muscles are probable late effects after growing pediatric patients receive proton CSI treatment [2]. The proposed method of early *in vivo* assessment of radiation damage has the potential to non-invasively confirm adequate vertebral body sparing while the patient is on treatment. Maintenance of healthy marrow bolsters the patient against the hematologic toxicities of systemic therapy and preserves patient candidacy for any additional treatment that may be required following radiation. Confirmation of vertebral body sparing further reduces the likelihood of vertebral growth inhibition, which can be a devastating side-effect in these patients with very favorable prognoses.

The current study only includes MR scans sampled between treatment fractions. Future investigations may focus on collecting comprehensive MR data for each treatment fraction to more accurately predict when bone marrow suppression occurs. The method presented here may further be integrated with a physics-informed deep learning-based CT material conversion method [18] to achieve vertebral body sparing intensity modulated proton therapy. Such a system would potentially improve treatment plan quality, verify treatment accuracy in real-time, and support medical decision-making for plan modification.

## 5 Conclusions

A method for the detection of early *in vivo* radiation damage is presented that uses serial MR scans to quantify vertebral marrow changes for pediatric CSI patients. The method can identify the transition from hematopoietic marrow to fatty marrow earlier than previously possible. Such a method can potentially quantify the quality of pediatric proton VBS CSI, which has the potential to spare pediatric patients from the most severe toxicities of this essential treatment.

**Acknowledgments**

This research is partly supported by the National Institutes of Health under Award Number R01CA215718, R01EB032680, R56EB033332 and P30CA138292.

**References**

[1] Ajithkumar T, Horan G, Padovani L, Thorp N, Timmermann B, Alapetite C, et al. SIOPE – Brain tumor group consensus guideline on craniospinal target volume delineation for high-precision radiotherapy. Radiotherapy and Oncology. 2018;128:192-7.

[2] Constine LS, Woolf PD, Cann D, Mick G, McCormick K, Raubertas RF, et al. Hypothalamic-Pituitary Dysfunction after Radiation for Brain Tumors. New England Journal of Medicine. 1993;328:87-94.

[3] Chang EL, Allen P, Wu C, Ater J, Kuttesch J, Maor MH. Acute toxicity and treatment interruption related to electron and photon craniospinal irradiation in pediatric patients treated at the University of Texas M. D. Anderson Cancer Center. International Journal of Radiation Oncology*Biology*Physics. 2002;52:1008-16.

[4] Packer RJ, Gajjar A, Vezina G, Rorke-Adams L, Burger PC, Robertson PL, et al. Phase III Study of Craniospinal Radiation Therapy Followed by Adjuvant Chemotherapy for Newly Diagnosed Average-Risk Medulloblastoma. Journal of Clinical Oncology. 2006;24:4202-8.

[5] Zhang R, Howell RM, Taddei PJ, Giebeler A, Mahajan A, Newhauser WD. A comparative study on the risks of radiogenic second cancers and cardiac mortality in a set of pediatric medulloblastoma patients treated with photon or proton craniospinal irradiation. Radiotherapy and Oncology. 2014;113:84-8.

[6] Paulino AC, Wen BC, Brown CK, Tannous R, Mayr NA, Zhen WK, et al. Late effects in children treated with radiation therapy for Wilms' tumor. International Journal of Radiation Oncology*Biology*Physics. 2000;46:1239-46.

[7] Paulino AC, Fowler BZ. Risk factors for scoliosis in children with neuroblastoma. International Journal of Radiation Oncology*Biology*Physics. 2005;61:865-9.

[8] De B, Florez MA, Ludmir EB, Maor MH, McGovern SL, McAleer MF, et al. Late Effects of Craniospinal Irradiation Using Electron Spinal Fields for Pediatric Cancer Patients. International Journal of Radiation Oncology*Biology*Physics. 2022.

[9] Hoeben BA, Carrie C, Timmermann B, Mandeville HC, Gandola L, Dieckmann K, et al. Management of vertebral radiotherapy dose in paediatric patients with cancer: consensus recommendations from the SIOPE radiotherapy working group. The Lancet Oncology. 2019;20:e155-e66.

[10] Paulino AC, Suzawa HS, Dreyer ZE, Hanania AN, Chintagumpala M, Okcu MF. Scoliosis in Children Treated With Photon Craniospinal Irradiation for Medulloblastoma. Int J Radiat Oncol Biol Phys. 2021;109:712-7.

[11] Weber DC, Habrand JL, Hoppe BS, Hill Kayser C, Laack NN, Langendijk JA, et al. Proton therapy for pediatric malignancies: Fact, figures and costs. A joint consensus statement from the pediatric subcommittee of PTCOG, PROS and EPTN. Radiotherapy and Oncology. 2018;128:44-55.

[12] Brown AP, Barney CL, Grosshans DR, McAleer MF, de Groot JF, Puduvalli VK, et al. Proton Beam Craniospinal Irradiation Reduces Acute Toxicity for Adults With Medulloblastoma. International Journal of Radiation Oncology*Biology*Physics. 2013;86:277-84.

[13] Eaton BR, Esiashvili N, Kim S, Weyman EA, Thornton LT, Mazewski C, et al. Clinical Outcomes Among Children With Standard-Risk Medulloblastoma Treated With Proton and Photon Radiation Therapy: A Comparison of Disease Control and Overall Survival. International Journal of Radiation Oncology*Biology*Physics. 2016;94:133-8.

[14] Farace P, Bizzocchi N, Righetto R, Fellin F, Fracchiolla F, Lorentini S, et al. Supine craniospinal irradiation in pediatric patients by proton pencil beam scanning. Radiotherapy and Oncology. 2017;123:112-8.

[15] Huynh M, Marcu LG, Giles E, Short M, Matthews D, Bezak E. Are further studies needed to justify the use of proton therapy for paediatric cancers of the central nervous system? A review of current evidence. Radiotherapy and Oncology. 2019;133:140-8.

[16] Paganetti H. Range uncertainties in proton therapy and the role of Monte Carlo simulations. Physics in Medicine and Biology. 2012;57:R99-R117.

[17] Chang C-W, Huang S, Harms J, Zhou J, Zhang R, Dhabaan A, et al. A standardized commissioning framework of Monte Carlo dose calculation algorithms for proton pencil beam scanning treatment planning systems. Medical Physics. 2020;47:1545-57.

[18] Chang C-W, Gao Y, Wang T, Lei Y, Wang Q, Pan S, et al. Dual-energy CT based mass density and relative stopping power estimation for proton therapy using physics-informed deep learning. Physics in Medicine & Biology. 2022;67:115010.

[19] Hajek PC, Baker LL, Goobar JE, Sartoris DJ, Hesselink JR, Haghighi P, et al. Focal fat deposition in axial bone marrow: MR characteristics. Radiology. 1987;162:245-9.

[20] Rosenthal D, Hayes C, Rosen B, Mayo-Smith W, Goodsitt M. Fatty Replacement of Spinal Bone Marrow due to Radiation: Demonstration by Dual Energy Quantitative CT and MR Imaging. J Comput Assist Tomogr. 1989;13(3):463-5.
[21] Krejcarek SC, Grant PE, Henson JW, Tarbell NJ, Yock TI. Physiologic and Radiographic Evidence of the Distal Edge of the Proton Beam in Craniospinal Irradiation. International Journal of Radiation Oncology*Biology*Physics. 2007;68:646-9.
[22] Gensheimer MF, Yock TI, Liebsch NJ, Sharp GC, Paganetti H, Madan N, et al. <em>In Vivo</em> Proton Beam Range Verification Using Spine MRI Changes. Int J Radiat Oncol Biol Phys. 2010;78:268-75.
[23] Blomlie V, Rofstad EK, Skjønsberg A, Tverå K, Lien HH. Female pelvic bone marrow: serial MR imaging before, during, and after radiation therapy. Radiology. 1995;194:537-43.
[24] Cavenagh EC, Weinberger E, Shaw DW, White KS, Geyer JR. Hematopoietic marrow regeneration in pediatric patients undergoing spinal irradiation: MR depiction. American Journal of Neuroradiology. 1995;16:461.
[25] ICRU78. Prescribing, Recording, and Reporting Proton-Beam Therapy. ICRU Publication 78. 2007.
[26] Relative Biological Effectiveness in Ion Beam Therapy. Vienna: INTERNATIONAL ATOMIC ENERGY AGENCY; 2008.
[27] Sonis AL, Tarbell N, Valachovic RW, Gelber R, Schwenn M, Sallan S. Dentofacial development in long-term survivors of acute lymphoblastic leukemia: A comparison of three treatment modalities. Cancer. 1990;66:2645-52.
[28] Willman KY, Cox RS, Donaldson SS. Radiation induced height impairment in pediatric Hodgkin's disease. International Journal of Radiation Oncology*Biology*Physics. 1994;28:85-92.
[29] Eifel PJ, Donaldson SS, Thomas PRM. Response of growing bone to irradiation: A proposed late effects scoring system. Int J Radiat Oncol Biol Phys. 1995;31:1301-7.
[30] Giantsoudi D, Seco J, Eaton BR, Simeone FJ, Kooy H, Yock TI, et al. Evaluating Intensity Modulated Proton Therapy Relative to Passive Scattering Proton Therapy for Increased Vertebral Column Sparing in Craniospinal Irradiation in Growing Pediatric Patients. Int J Radiat Oncol Biol Phys. 2017;98:37-46.
[31] Yankelevitz DF, Henschke CI, Knapp PH, Nisce L, Yi Y, Cahill P. Effect of radiation therapy on thoracic and lumbar bone marrow: evaluation with MR imaging. American Journal of Roentgenology. 1991;157:87-92.